\documentclass[12pt,twoside]{article}
\usepackage{fleqn,espcrc1,epsfig}
\usepackage{graphicx}
\usepackage[figuresright]{rotating}

\begin{document}

\title{Semiempirical Shell Model Masses with Magic Number ${\bf Z =
126}$ for Translead Elements with ${\bf N \leq 126}$}
\author{S. Liran\footnote{Present address: Kashtan 3/3, Haifa 34984,
Israel}, A. Marinov and N. Zeldes \\
The Racah Institute of Physics, The Hebrew University of Jerusalem, Jerusalem 91904,
Israel}
\maketitle

\begin{abstract}
A semiempirical shell model mass equation based on magic number $Z =
126$ and applicable to translead elements with $N \leq 126$ is
presented. For $\alpha$-decay energies the equation is shown to have a
high predictive power and an rms deviation from the data of about 100
keV. The rms deviations for masses and other mass differences are between
about 200 and 300 keV.
\end{abstract}

\bigskip
PACS numbers: 21.10.Dr, 21.60.Cs, 27.80.+w

\bigskip

Recent progress in superheavy elements (SHE) research reaching to
$^{293}118$ and its $\alpha$-decay products \cite{nin99} makes it
necessary to find an appropriate substitute for the semiempirical
shell-model mass equation (SSME) \cite{liz76} (see also ref.
\cite{zel96}) for nuclei in the neighbourhood of $Z=114$ and beyond
\cite{limaz}. The $\alpha$ energies of the decaying chain vary
smoothly from $^{293}118$ to $^{269}$Sg $(Z=106)$, with no indication
of magicity at $Z=114$ in these nuclei \cite{nin99}, whereas the SSME
assumes that $Z=114$ is the next spherical proton magic number after lead and
it stops there. Furthermore, the SSME becomes unsuitable for
extrapolation already earlier, beyond Hs $(Z=108)$, as shown by its
increasing deviations from the data when $Z$ increases. (Like in fig.
4 of ref. \cite{nin99}.)

Recent phenomenological studies of BE(2) systematics \cite{zam95} and
of the persistence of the Wigner term in masses of heavy nuclei
\cite{zel98} indicate $Z=126$ as the next spherical proton magic number
after lead, and this is consistent with considerations based on
nuclear diffuseness \cite{mys98}. Recent self-consistent and
relativistic mean-field calculations \cite{cwi96,lal96,rut97,ben99,kru00}
variously predict proton magicities for $Z=114,120,124$ and 126.

During the early stages of the SSME \cite{liran}, when it was adjusted
separately in individual shell regions in the $N-Z$ plane, both $Z=114$
and $Z=126$, which were at the time considered possible candidates
for the postlead proton magic number (see, e.g., ref. \cite{kum89}),
were tried as a shell region boundary in each of the two
heaviest regions with $Z \geq 82$ and respective $N$-boundaries $82 \leq
N \leq 126$ (called here region A) and $126 \leq N \leq 184$ (called
region B). The agreement with the data was about the same for both
choices, and the prevailing view in the mid nineteen-seventies led to
the choice of $Z=114$ for the SSME mass table \cite{liz76}. In a
recent communication \cite{limaz} we showed that the early $Z=126$
results have a high predictive power in the interior of region B and
proposed their use there as a predictive tool in SHE research. In the
present note we study the predictive power or extrapolatability of the
early $Z=126$ results in region A and propose a mass equation based on
them as a substitute for the SSME \cite{liz76} in the interior of the
region. The study is based on the newer data that became available
after the adjustments were made, as was done in refs.
\cite{limaz,hau84,mol95,mnk97,abo95}.

In the SSME the total nuclear energy $E$ is a sum of pairing, deformation and
Coulomb energies:

\begin{equation}
E\left( {N,Z} \right)=E_{pair}\left( {N,Z} \right)+E_{def}\left(
{N,Z} \right) +E_{Coul}\left( {N,Z} \right) \ .
    \label{eq1}
\end{equation}
The form of $E_{Coul}$ is the same in all shell regions:

\begin{equation}
E_{Coul}\left( {N,Z} \right)=\left( {{{2Z_0} \over A}}
\right)^{{\raise3pt\hbox{$\scriptstyle 1$} \!\mathord{\left/
{\vphantom {\scriptstyle {1 3}}}\right.\kern-\nulldelimiterspace}
\!\lower3pt\hbox{$\scriptstyle 3$}}}[ {\alpha ^c+\beta ^c\left(
{Z-Z_0} \right)+\gamma ^c\left( {Z-Z_0} \right)^2} ]\  ,
    \label{eq2}
\end{equation}
and that of $E_{pair}$ is the same separately in all diagonal shell
regions, where the major valence shells are the same for neutrons and
for protons, and in all non-diagonal regions, where the neutron and
proton valence shells are different. Unlike in \cite{liz76}, with
$Z=126$ rather than 114 as an upper proton boundary, region A becomes
a diagonal region with

\begin{eqnarray}
& & E_{pair}\left( {N,Z} \right) = \nonumber \\
& & \left( {{{A_0} \over A}} \right)\left[ {\alpha + \beta \left( {A-A_0}
  \right)+\gamma \left( {A-A_0} \right)^2+\varepsilon T\left( {T+1}
  \right)+{{1-\left( {-1} \right)^A} \over 2}\Theta +{{1-\left( {-1}
  \right)^{NZ}} \over 2}\kappa } \right] \ .
\label{eq3}
\end{eqnarray}
The part $E_{def}$ for region A with $Z=126$ as upper proton boundary
is \cite{liran}

\begin{equation}
E_{def}\left( {N,Z} \right)=\left( {{{A_0} \over A}} \right)\left[
{\varphi _{11}\Phi _{11}\left( {N,Z} \right)+\psi _{20}\left[
{\Psi _{20}\left( {N,Z} \right)+\Psi _{20}\left( {Z,N} \right)} \right]}
\right]
    \label{eq4}
\end{equation}
with
\begin{equation}
\Phi _{11}\left( {N,Z} \right)=\left( {N-82} \right)\left(
{126-N} \right)\left( {Z-82} \right)\left( {126-Z} \right)
    \label{eq5}
\end{equation}
\begin{equation}
\Psi _{20}\left( {N,Z} \right)=\left( {N-82} \right)^2\left( {126-N}
\right)^2\left( {N-104} \right) \ .
    \label{eq6}
\end{equation}
In eqs.\ (\ref{eq2})$-$(\ref{eq4}) $A=N+Z$ and $T = |T_{z}| = {1 \over 2}
|N - Z|$\footnote{In the as yet unknown odd-odd $N = Z$ translead nuclei the
g.s. is expected to have $T = |T_{z}|+1$ and seniority zero, whereas
eq.\ (\ref{eq1}) with $T = |T_{z}|$ gives the energy of a low
excited seniority two state \cite{zel96}. (See also ref. \cite{zel99}.)}.
The respective values of $Z_{0}$ and $A_{0}$ are 82 and 164. The
coefficients multiplying the functions of $N$ and $Z$ are adjustable
parameters which were determined by a least-squares adjustment to the
data \cite{liran}. Their numerical values resulting from that
adjustment are given in the second column of table I. The mass
excesses $\Delta M(N,Z)$ are obtained by adding to eq.\ (\ref{eq1}) the
sum of nucleon mass excesses $N\Delta M_{n}+Z\Delta M_{H}$.

The experimental data used in the adjustment included 29 masses and
62 $Q_{\alpha}$ values connecting unknown masses (Ref. \cite{wag71}
augmented by data from the literature up to Spring 1973). Presently
there are 150 known experimental masses in region A and 3
$Q_{\alpha}$ values connecting unknown masses (Refs. \cite{auw95}
(excluding values denoted ``systematics'' (\#)) and \cite{radon}, augmented
by data from the recent literature). There are 121 new masses that were not
used in the adjustments.

Fig. 1 shows the deviations from the data of the predictions of eq.\
(\ref{eq1}) (with the definitions (\ref{eq2})$-$(\ref{eq4}) and the
coefficients from the second column of table I) for all the 150 known masses.
The deviations are plotted as function of the distance from the line
of $\beta$-stability, denoted ``neutrons from stability'' ($NFS$) and
defined by $NFS = N - Z - {{0.4A^{2}}\over{A+200}}$ \cite{hau84}.
Empty circles denote the deviations of the 29 originally adjusted
masses and full circles mark the deviations of the 121 new masses.
The original deviations are relatively small. The new deviations
are mostly considerably larger and are almost all negative.
There is a large scatter of the points, superposed on an overall
decreasing trend when NFS increases in the negative direction. For closer
scrutiny we show in fig. 2 lines connecting deviations of Pb, Bi, Po and At
isotopes as function of $T$ rather than $NFS$. Isotopic lines of other
elements follow closely. There is an odd-even staggering
of the points, where for even-even and odd-odd nuclei the deviations
are respectively higher and lower than for odd-mass ones, and there is
an overall increasing overbinding when $T$ decreases away from
stability. The increasing staggering when $T$ decreases as compared to
the old data indicates increasing pairing parameters $\Theta$ (see also ref.
\cite{jhj84}) and (even more so) $\kappa$ towards the proton drip line.
Similarly, the increasing binding when $T$ decreases indicates that the
symmetry-energy parameter $\varepsilon$ decreases towards the drip line.

Table II, patterned after similar more elaborate ones
\cite{mol95,mnk97} (see also ref. \cite{limaz}), shows the values of $\delta_{av}$ and $\delta_{rms}$, the
respective average and rms deviations of eq.\ (\ref{eq1}) from the
data, for $\Delta M, S_{n}, S_{p}, Q_{\beta^{-}}$ and $Q_{\alpha}$.
The deviations are shown separately for the older data that were used
in the adjustment and for the newer data. The last column shows the
error ratios $\delta^{new}_{rms}:\delta^{old}_{rms}$.

For the older data the $\delta_{av}$ are few tens keV at most, and
the $\delta_{rms}$ are in the range 100-250 keV. For the newer nuclei
the agreement with the data of predicted $\Delta M$ (fig. 1), $S_{n}, S_{p}$ and
$Q_{\beta^{-}}$ values has much deteriorated. On the other hand,
the $\delta_{rms}$ of the predicted $Q_{\alpha}$ values has even very
slightly improved. The $Q_{\alpha}$ deviations, both old and new, are
remarkably small.

The high degree of extrapolatability of the $Q_{\alpha}$ values as compared
to the poorer predictions of $\Delta M, S_{n}, S_{p}$ and
$Q_{\beta^{-}}$ is presumably due to the composition of the old data
set used in the adjustments. The coefficients $\beta, \gamma,
\beta^{c}, \gamma^{c}, \varphi_{11}$ and $\psi_{20}$,
which contribute to both mass and $Q_{\alpha}$ values, and also
$\alpha^{c}$, which was determined using $\beta^{c}$ and $\gamma^{c}$
\cite{liran} (see also ref. \cite{ash90}), were determined
by all the 91 available mass and $Q_{\alpha}$ data.
On the other hand, neglecting their $A$-dependence the coefficients
$\alpha, \varepsilon, \Theta$ and $\kappa$ cancel in $Q_{\alpha}$ and
they were determined essentially by the smaller group of 29
masses found in the nearest-to-stability corner of the shell region,
with too-small values of $\Theta$ and $\kappa$ and too large value of
$\varepsilon$ as compared to more proton-rich nuclei. These values,
which to a large extent cancel in $Q_{\alpha}$, are responsible for
the large deviations of the predicted new masses.

One would like to restore to the new $\Delta M, S_{n}, S_{p}$ and
$Q_{\beta^{-}}$ predictions the same degree of agreement with the data
as the old predictions had, while at the same time retaining the high
quality of $Q_{\alpha}$ predictions. For eq.\ (\ref{eq1}) this goal
might most simply be
approached by making a least-squares adjustment of eq.\ (\ref{eq1})
to all the 150 known masses, with only four adjustable parameters
$\alpha, \varepsilon, \Theta$ and $\kappa$, while the other seven
coefficients are held fixed on their old values from the second column
of table I. The re-adjusted parameters $\alpha$ and $\varepsilon$ would correct
the systematic overbinding when $T$ decreases, and the re-adjusted
$\Theta$ and $\kappa$ would decrease the odd-even staggering of the
deviations.

The re-adjusted values of the coefficients $\alpha, \varepsilon, \Theta$
and $\kappa$ are given in the third column of table I. Their changes
compared to the second column are in the anticipated directions. The
respective average and rms mass deviations obtained in the adjustment
for all the 150 mass data are 2 and 246 keV.

Fig. 3 shows the
deviations from the data of the predicted mass values resulting from
the new adjustment, similarly to fig. 1. For ease of comparison the
same nuclei are denoted by the same kind of circle (empty or full) in
the two figures. The deviations for the post 1973 measured nuclei
(denoted by full circles) and their odd-even staggering are
considerably smaller than in fig. 1, and the overall decreasing trend
when $T$ decreases has largely disappeared, as was
expected. The deviations for the older data have worsened, though, and
there is a new overall oscillatory trend not
observed in fig. 1, indicating that enlarging eq.\ (\ref{eq1}) by additional
particle-hole symmetric $E_{def}$ terms (eq.\ (\ref{eq4}))
might reduce the deviations further.

Table III shows the resulting $\delta_{av}$ and $\delta_{rms}$ values
of the predicted deviations. Like in table II they are shown
separately for the old 1973 data nearer to stability and for the new
data extending into the interior of region A towards the proton drip
line. For $S_{n}, S_{p}$ and $Q_{\alpha}$ the new deviations
are similar to those of the original nuclei in table II.
For $\Delta M$ and $Q_{\beta^{-}}$ the $\delta_{rms}$ are
respectively 1.5 and 1.1 times larger.

The new mass deviations shown in table III are about one half to two thirds of
the corresponding deviations of several recent mass models
\cite{mol95,mnk97,abo95,mys96,duz95}, and for $Q_{\alpha}$ they are
smaller. The smaller values of the deviations in table III are
presumably due mainly to the inclusion of the particle-hole symmetric
configuration-interaction terms $E_{def}$ (eq.\ (\ref{eq4})) in eq.\
(\ref{eq1}) \cite{limaz}. Until a new adjustment of the $Z = 126$
SSME to the data in both regions A and B is undertaken, which would
further reduce the deviations in both regions, we propose the
use of eq.\ (\ref{eq1}), with the new values of $\alpha, \varepsilon,
\Theta$ and $\kappa$ and the old values of the other coefficients, given
in table I, as a predictive tool for masses and their differences , and
particularly for the extrapolatable-proven
$Q_{\alpha}$ values, in the interior of region A.

It should be emphasized, though, that the above results are not a
proof of superior magicity of $Z=126$ in region A as compared to other
recently proposed prediction \cite{rut97,ben99,kru00}, because no
comparative studies of this kind were made. (See also ref.
\cite{limaz}.)

We thank the Atomic Mass Data Center in Orsay for ref. \cite{radon} and Stelian
Gelberg and Dietmar Kolb for help with the calculations.

\newpage

%\end{document}

%\documentstyle[12pt,aps,prc,preprint]{revtex}
%\begin{document}

\begin{table}
    \caption{Values of the coefficients of eq. (1) as
    determined by adjustment to the old data [13] (Old Value) and
    as readjusted in the present work (New Value).}
\vspace*{0.4cm}
    \begin{tabular}{ccc}
        coefficient & Old Value (keV) & New Value (keV) \\
        \hline
        $\alpha$ & $-1.9902575 \times 10^{6}$ & $-1.987628 \times 10^{6}$  \\

        $\beta$ & $-2.4773664 \times 10^{4}$ &   \\

        $\gamma$ & $-8.51085 \times 10^{1}$ &   \\

        $\varepsilon$ & $4.658516 \times 10^{2}$ & $4.585496 \times 10^{2}$  \\

        $\Theta$ & $9.762 \times 10^{2}$ & $1.2183 \times 10^{3}$  \\

        $\kappa$ & $1.4965 \times 10^{3}$ & $2.1937 \times 10^{3}$  \\

        $\alpha^{c}$ & $7.968418 \times 10^{5}$ &   \\

        $\beta^{c}$ & $2.032906 \times 10^{4}$ &   \\

        $\gamma^{c}$ & $9.819137 \times 10^{1}$ &   \\

        $\varphi_{11}$ & $-4.794 \times 10^{-2}$ &   \\

        $\psi_{20}$ & $9.095 \times 10^{-4}$ &   \\
    \end{tabular}
\end{table}

\begin{table}
    \caption{Numbers of data N, average deviations $\delta_{av}$, and
    rms deviations $\delta_{rms}$, for eq.~(1) with the old values of the
    coefficients from table I. The last column shows the error ratios
    $\delta^{new}_{rms}:\delta^{old}_{rms}$.}

    \vspace*{0.4cm}
    \begin{tabular}{lcccccccc}
    & \multicolumn{3}{c}{Original nuclei (1973)} & &
    \multicolumn{3}{c}{New nuclei (1973-2000)}  \\
         &  & $\delta_{av}$ & $\delta_{rms}$ & & & $\delta_{av}$ &
         $\delta_{rms}$ & Error  \\
        Data & N & (keV) & (keV) & & N & (keV) & (keV) & ratio \\
        \hline
        $\Delta M$ & 29 & $-29$ & 146 & & 121 & $-807$ & 1008 & 6.88 \\
        $S_{n}$ & 18 & 39 & 214 &  & 120 & $-120$ & 406 & 1.90 \\
        $S_{p}$ & 22 & 9 & 182 &  & 104 & 132 & 417 & 2.29 \\
        $Q_{\beta^{-}}$ & 15 & $-20$ & 242 &  & 101 & 248 & 583 & 2.41 \\
        $Q_{\alpha}$ & 78 & 5 & 103 &  & 31 & 40 & 89 & 0.87 \\
    \end{tabular}
\end{table}

\begin{table}
    \caption{Numbers of data N, average deviations $\delta_{av}$, and
    rms deviations $\delta_{rms}$, for eq.~(1) with the new values of the
    coefficients $\alpha, \varepsilon, \Theta, \kappa$ and the old values
    of the other seven coefficients from Table I.}

    \vspace*{0.4cm}
    \begin{tabular}{lccccccc}
    & \multicolumn{3}{c}{Original nuclei (1973)} & &
    \multicolumn{3}{c}{New nuclei (1973-2000)}  \\
         &  & $\delta_{av}$ & $\delta_{rms}$ & & & $\delta_{av}$ &
         $\delta_{rms}$  \\
        Data & N & (keV) & (keV) & & N & (keV) & (keV) \\
        \hline
        $\Delta M$ & 29 & $-193$ & 344 & & 121 & $48$ & 216  \\
        $S_{n}$ & 18 & 158 & 416 &  & 120 & $-10$ & 205  \\
        $S_{p}$ & 22 & $-144$ & 202 &  & 104 & 18 & 184  \\
        $Q_{\beta^{-}}$ & 15 & $-257$ & 475 &  & 101 & 15 & 277  \\
        $Q_{\alpha}$ & 78 & $-5$ & 104 &  & 31 & 18 & 85  \\
    \end{tabular}
\end{table}

\begin{figure}[h]
\includegraphics[width=1.0\textwidth]{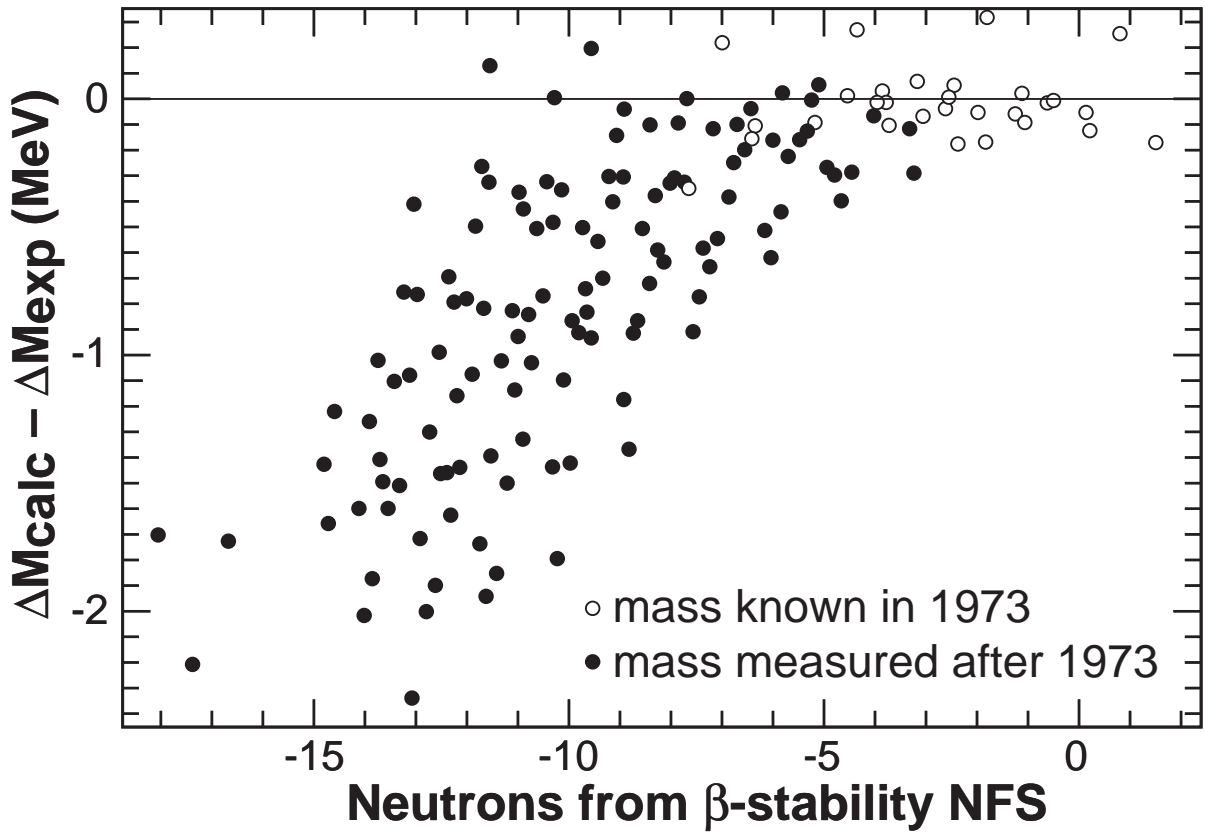}
\vspace*{-0.9cm} \caption{Deviations of the predicted masses [13]
    from the presently known data in region A. Shown as function of
    Neutrons From Stability $(NFS)$.}
\end{figure}

\begin{figure}[h]
\begin{center}
\includegraphics[width=0.5\textwidth]{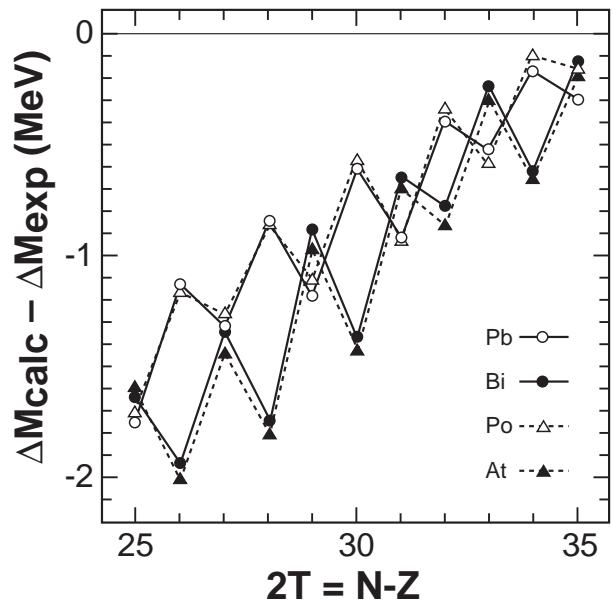}
\vspace*{-0.9cm} \caption{Isotopic lines of predicted mass
deviations [13] of Pb, Bi, Po and At nuclei measured after the
    adjustments were made. Shown as function of $T$.}
\end{center}
\end{figure}

\begin{figure}[h]
\includegraphics[width=1.0\textwidth]{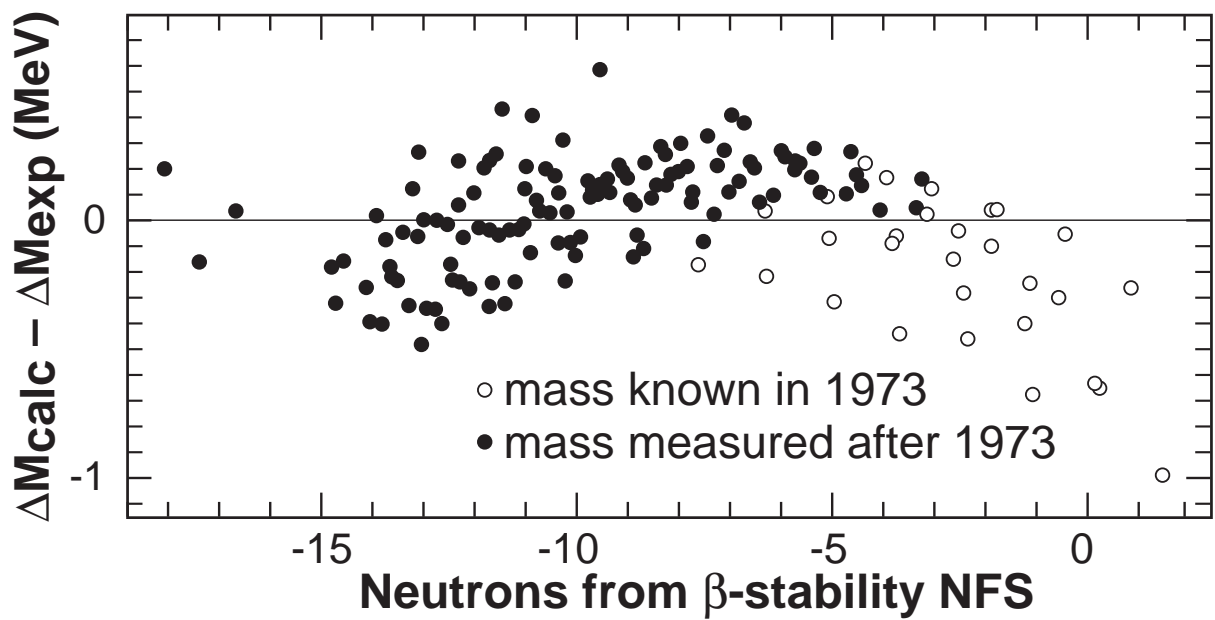}
\vspace*{-0.9cm} \caption{Deviations of the predicted masses
resulting from
    the present adjustment from the presently known data in region A.
    Shown as function of $NFS$.}
\end{figure}

\end{document}